\documentclass[preprint1]{aastex}
\usepackage{amsmath}
\slugcomment{}

\newcommand{\be}{\begin{equation}}
\newcommand{\ee}{\end{equation}}

\shorttitle{Old Passive Galaxies}
\shortauthors{Wei, Wu, \& Melia}

\begin{document}

\title{The Age-Redshift Relationship of Old Passive Galaxies}
\author{Jun-Jie Wei\altaffilmark{1,2}, Xue-Feng Wu\altaffilmark{1,3,4},
Fulvio Melia\altaffilmark{1,5}, Fa-Yin Wang\altaffilmark{6,7}, and Hai Yu\altaffilmark{6,7}}
\altaffiltext{1}{Purple Mountain Observatory, Chinese Academy of Sciences, Nanjing 210008,
China; jjwei@pmo.ac.cn, xfwu@pmo.ac.cn.}
\altaffiltext{2}{University of Chinese Academy of Sciences, Beijing 100049, China.}
\altaffiltext{3}{Chinese Center for Antarctic Astronomy, Nanjing 210008, China.}
\altaffiltext{4}{Joint Center for Particle, Nuclear Physics and Cosmology, Nanjing
University-Purple Mountain Observatory, Nanjing 210008, China.}
\altaffiltext{5}{Department of Physics, The Applied Math Program, and Department of Astronomy,
The University of Arizona, AZ 85721, USA; fmelia@email.arizona.edu.}
\altaffiltext{6}{School of Astronomy and Space Science, Nanjing University, Nanjing 210093, China.}
\altaffiltext{7}{Key Laboratory of Modern Astronomy and Astrophysics (Nanjing University), Ministry of Education, Nanjing 210093, China.}

\begin{abstract}
We use 32 age measurements of passively evolving galaxies as a function of redshift
to test and compare the standard model ($\Lambda$CDM) with the
$R_{\rm h}=ct$ Universe. We show that the latter fits the data with
a reduced $\chi^2_{\rm dof}=0.435$ for a Hubble constant $H_{0}=
67.2_{-4.0}^{+4.5}$ km $\rm s^{-1}$ $\rm Mpc^{-1}$. By comparison, the optimal
flat $\Lambda$CDM model, with two free parameters (including
$\Omega_{\rm m}=0.12_{-0.11}^{+0.54}$ and $H_{0}=94.3_{-35.8}^{+32.7}$ km $\rm s^{-1}$
$\rm Mpc^{-1}$), fits the age-\emph{z} data with a reduced
$\chi^2_{\rm dof}=0.428$. Based solely on their $\chi^2_{\rm dof}$
values, both models appear to account for the data very well, though
the optimized $\Lambda$CDM parameters are only marginally consistent with
those of the concordance model ($\Omega_{\rm m}=0.27$ and $H_{0}=
70$ km $\rm s^{-1}$ $\rm Mpc^{-1}$). Fitting the age-$z$ data with the
latter results in a reduced $\chi^2_{\rm dof}=0.523$.
However, because of the different number of free parameters
in these models, selection tools, such as the Akaike, Kullback
and Bayes Information Criteria, favour $R_{\rm h}=ct$ over $\Lambda$CDM
with a likelihood of $\sim 66.5\%-80.5\%$ versus $\sim 19.5\%-33.5\%$.
These results are suggestive, though not yet compelling, given the
current limited galaxy age-$z$ sample. We carry out Monte Carlo
simulations based on these current age measurements to
estimate how large the sample would have to be in
order to rule out either model at a $\sim 99.7\%$ confidence level. We
find that if the real cosmology is $\Lambda$CDM, a sample of $\sim 45$
galaxy ages would be sufficient to rule out $R_{\rm h}=ct$ at this level of
accuracy, while $\sim 350$ galaxy ages would be required to rule out
$\Lambda$CDM if the real Universe were instead $R_{\rm h}=ct$. This difference
in required sample size reflects the greater number of free parameters available
to fit the data with $\Lambda$CDM.
\end{abstract}


\section{Introduction\label{sec:intro}}
In recent years, the cosmic evolution has been studied using a diversity of
observational data, including cosmic chronometers (Melia \& Maier 2013),
Gamma-ray bursts (GRBs; Wei et al. 2013), high-\emph{z} quasars (Melia 2013,
2014a), strong gravitational lenses (Wei et al. 2014; Melia et al. 2015),
and Type Ia SNe (Wei et al. 2015). In particular, the predictions of $\Lambda$CDM
have been compared with those of a cosmology we refer to as the $R_{\rm h}=ct$
Universe (Melia 2007; Melia \& Shevchuk 2012). In all such one-on-one comparisons
completed thus far, model selection tools show that the data favour $R_{\rm h}=ct$
over $\Lambda$CDM.

The $R_{\rm h}=ct$ Universe is a Friedmann-Robertson-Walker (FRW) cosmology
that has much in common with $\Lambda$CDM, but includes an additional
ingredient motivated by several theoretical and observational arguments
(Melia 2007; Melia \& Abdelqader 2009; Melia \& Shevchuk 2012; see also
Melia 2012a for a more pedagogical treatment). Like $\Lambda$CDM, it
adopts an equation of state $p=w\rho$, with $p=p_{\rm m}+p_{\rm r}+
p_{\rm de}$ (for matter, radiation, and dark energy, respectively)
and $\rho=\rho_{\rm m}+\rho_{\rm r}+\rho_{\rm de}$, but
goes one step further by specifying that $w=(\rho_{\rm r}/3+w_{\rm de}
\rho_{\rm de})/\rho=-1/3$ at all times. Here, $p$ is the total pressure and
$\rho$ is the total energy density. One might come away with the
impression that this equation of state cannot be consistent with that
(i.e., $w=[\rho_{\rm r}/3-\rho_\Lambda]/\rho$) in the standard model. But
in fact if we ignore the constraint $w=-1/3$ and
instead proceed to optimize the parameters in $\Lambda$CDM by fitting
the data, the resultant value of $w$ averaged over a Hubble time is
actually $-1/3$ within the measurement errors (Melia 2007; Melia \&
Shevchuk 2012). In other words, though $w=(\rho_{\rm r}/3-\rho_\Lambda)/\rho$
in $\Lambda$CDM may be different from $-1/3$ from one moment to the next, its
value averaged over the age of the Universe equals what it would have
been in $R_{\rm h}=ct$ all along (Melia 2015).

In this paper, we continue to compare the predictions of $\Lambda$CDM with
those in the $R_{\rm  h}=ct$ Universe, this time focusing on the age-redshift
relationship, which differs from one expansion scenario to another. Though
the current age of the universe may be similar in these two cosmologies,
the age versus redshift relationship is not, particularly at high redshifts
(Melia 2013, 2014b). Previous work with the age estimates of distant
objects has already provided effective constraints on cosmological parameters
(see, e.g., Alcaniz \& Lima 1999; Lima \& Alcaniz 2000; Jimenez \& Loeb 2002;
Jimenez et al. 2003; Capozziello et al. 2004; Fria{\c c}a et al. 2005; Simon et
al. 2005; Jain \& Dev 2006; Pires et al. 2006; Dantas et al. 2007, 2009, 2011;
Samushia et al. 2010). For example, using the simple criterion that the age
of the Universe at any given redshift should always be greater than or
equal to the age of the oldest object(s) at that redshift, the measured
ages were used to constrain parameters in the standard
model (Alcaniz \& Lima 1999; Lima \& Alcaniz 2000; Jain \& Dev 2006;
Pires et al. 2006; Dantas et al. 2007, 2011).  Cosmological parameters
have also been constrained by the measurement of the differential age
$\Delta z/\Delta t$, where $\Delta z$ is the redshift separation between
two passively evolving galaxies having an age difference $\Delta t$ (Jimenez
\& Loeb 2002; Jimenez et al. 2003). And the lookback time versus redshift
measurements for galaxy clusters and passively evolving galaxies have
been used to constrain dark energy models (Capozziello et al. 2004;
Simon et al. 2005; Dantas et al. 2009; Samushia et al. 2010).
This kind of analysis is therefore particularly
interesting and complementary to those mentioned earlier, which are
essentially based on distance measurements to a particular class of
objects or physical rulers (see Jimenez \& Loeb 2002, for a discussion
on cosmological tests based on relative galaxy ages).

Age measurements of high-$z$ objects have been valuable in constraining
the cosmological parameters of the standard model even before dark energy
was recognized as an essential component of the cosmic fluid
(see, e.g., Bolte \& Hogan 1995; Krauss \& Turner 1995; Dunlop et al. 1996;
Alcaniz \& Lima 1999; Jimenez \& Loeb 2002). Of direct relevance to the
principal aim of this paper is the fact that, although the distance--redshift
relationship is very similar in $\Lambda$CDM and the
$R_{\rm h}=ct$ Universe (even out to $z\ga 6-7$; Melia 2012b; Wei et al. 2013),
the age--redshift dependence is not. This difference is especially noticeable
in how we interpret the formation of structure in the early Universe. For
example, the emergence of quasars at $z\ga 6$, which are now known to
be accreting at, or near, their Eddington limit (see, e.g., Willott et~al.
2010; De~Rosa et~al. 2011). This presents a problem for $\Lambda$CDM because
it is difficult to understand how $\sim 10^9\;M_\odot$ supermassive black
holes could have appeared only 700--900 Myr after the big bang. Instead, in
$R_{\rm h}=ct$, their emergence at redshift $\sim 6$ corresponds to a
cosmic age of $\ga 1.6$ Gyr, which was enough time for them to begin
growing from $\sim 5-20\; M_\odot$ seeds (presumably the remnants of
Pop~II and~III supernovae) at $z\la 15$ (i.e., \emph{after} the onset
of re-ionization) and still reach a billion solar masses by $z\sim 6$
via standard, Eddington-limited accretion (Melia 2013).

In this paper, we will broaden the base of support for this cosmic probe
by demonstrating its usefulness in testing {\it competing} cosmological models.
Following the methodology presented in Dantas et al. (2011), we will use 32 age
measurements of passively evolving galaxies as a function of redshift
(in the range $0.117\leq z \leq1.845$) to test the predicted age-redshift
relationship of each model. From an observational viewpoint, because
the age of a galaxy must be younger than the age of the Universe at any
given redshift, there must be an incubation time, or delay factor $\tau$,
for the galaxy to form after the big bang. In principle, there could be a
different $\tau_{i}$ for each object $i$ since galaxies can form at different epochs.
However, the simplest approach we can take is to begin with the assumption
made in earlier work (see, e.g., Dantas et al. 2009, 2011; Samushia et al. 2010),
i.e., we will adopt an average delay factor $\langle\tau\rangle$ and
use it uniformally for every galaxy. But we shall also consider cases in
which the delay factors $\tau_{i}$ are distributed, and study the impact
of this non-uniformity on the overall fits to the data.

We will demonstrate that the current sample of galaxy ages favours the
$R_{\rm h}=ct$ Universe with a likelihood of $\sim 66.5-80.5\%$ of
being correct, versus $\sim 19.5-33.5\%$ for $\Lambda$CDM.
Though this result is still only marginal, it nonetheless calls for a
significant increase in the sample of suitable galaxy ages in order
to carry out more sophisticated and higher precision measurements.
We will therefore also construct mock catalogs to investigate how
big the sample of measured galaxy ages has to be in order to
rule out one (or more) of these models.

The outline of this paper is as follows. In \S~2, we will briefly describe
the age-redshift test, and then constrain the cosmological parameters---both
in the context of $\Lambda$CDM and the $R_{\rm h}=ct$ universe, first
using a uniform (average) delay factor $\langle\tau\rangle$ (\S~3), and then
a distribution of $\tau_i$ values (\S~4). In \S~5, we will discuss the model
selection tools we use to test $\Lambda$CDM and the $R_{\rm h}=ct$
cosmologies. In \S~6, we will estimate the sample size required from future
age measurements to reach likelihoods of $\sim 99.7\%$ and $\sim0.3\%$
(i.e., $3\sigma$ confidence limits) when using model selection tools to
compare these two models, and we will end with our conclusions in \S~7.

\section{The age-redshift test\label{sec:intro}}
The theoretical age of an object at redshift $z$ is given as
\begin{equation}
t^{\rm th}(z,\mathbf{p})=\int_{z}^{\infty}\frac{dz'}{(1+z')H(z',\mathbf{p})}\;,
\end{equation}
where $\mathbf{p}$ stands for all the parameters of the cosmological model
under consideration and $H(z,\mathbf{p})$ is the Hubble parameter at redshift
\emph{z}. From an observational viewpoint, the total age of a given
object (e.g., a galaxy) at redshift \emph{z} is given by $t^{\rm obs}(z)=
t_{G}(z)+\tau$, where $t_{G}(z)$ is the estimated age of its oldest stellar
population and $\tau$ is the incubation time, or delay factor, which accounts
for our ignorance about the amount of time elapsed since the big bang
to the epoch of star creation.

To compute model predictions for the age $t^{\rm th}(z,\mathbf{p})$ in Equation~(1),
we need an expression for $H(z,\mathbf{p})$. As we have seen, $\Lambda$CDM assumes
specific constituents in the density, written as $\rho=\rho_{\rm r}+\rho_{\rm m}+
\rho_{\rm de}$. These densities are often written in terms of today's critical
density, $\rho_c\equiv 3c^2 H_0^2/8\pi G$, represented as $\Omega_{\rm m}\equiv
\rho_{\rm m}/\rho_c$, $\Omega_{\rm r}\equiv\rho_{\rm r}/\rho_c$, and
$\Omega_{\rm de}\equiv \rho_{\rm de}/\rho_c$. $H_0$ is the Hubble constant.
In a flat universe with zero spatial curvature, the total scaled energy
density is $\Omega\equiv\Omega_{\rm m}+\Omega_{\rm r}+\Omega_{\rm de}=1$.
When dark energy is included with an unknown equation-of-state,
$p_{\rm de}=w_{\rm de}\rho_{\rm de}$, the most general expression for
the Hubble parameter is
\begin{equation}
H(z,\mathbf{p})=H_{0}\left[\Omega_m(1+z)^{3}+
\Omega_{\rm r}(1+z)^{4}+\Omega_k(1+z)^{2}+\Omega_{\rm de}(1+z)^{3(1+w_{\rm de})}\right]^{1/2},
\end{equation}
where $\Omega_k$ is defined similarly to $\Omega_{\rm m}$ and represents the spatial curvature
of the Universe. Of course, $\Omega_{\rm r}$ ($\sim5\times10^{-5}$) is known from the current temperature
($\simeq2.725$ K) of the CMB, and the value of $H_0$. In addition, we assume a flat
$\Lambda$CDM cosmology with $\Omega_k=0$, for which $\Omega_{\rm de}=1-\Omega_{\rm m}
-\Omega_{\rm r}$, thus avoiding the introduction of $\Omega_{\rm de}$ as an additional
free parameter. So for the basic $\Lambda$CDM model, $\mathbf{p}$ includes three
free parameters: $\Omega_{\rm m}$, $w_{\rm de}$, and $H_{0}$, though to be as favorable
as possible to this model, we will also assume that dark energy is a cosmological
constant, with $w_{\rm de}=-1$, leaving only two adjustable parameters. However, when
we consider the concordance model with the assumption of prior parameter values, the
fits have zero degrees of freedom from the cosmology itself.

The $R_{\rm h}=ct$ Universe is a flat Friedmann-Robertson-Walker (FRW)
cosmology that strictly adheres to the constraints imposed by the
simultaneous application of the cosmological principle and Weyl's
postulate (Melia 2012b; Melia \& Shevchuk 2012). When these ingredients
are applied to the cosmological expansion, the gravitational horizon
$R_{\rm h}=c/H$ must always be equal to $ct$. This cosmology is therefore
very simple, because $a(t)\propto t$, which also means that $1 + z = 1/t$,
with the (standard) normalization that $a(t_{0}) = 1$. Therefore,
in the $R_{\rm h}=ct$ Universe, we have the straightforward scaling
\begin{equation}
H(z,\mathbf{p})=(1+z)H_0.
\end{equation}
Notice, in particular, that the expansion rate $H(z)$ in this model has
only one free parameter, i.e., $\mathbf{p}$ is $H_{0}$. From Equation~(3),
the age of the $R_{\rm h}=ct$ Universe at redshift $z$ is simply
\begin{equation}
t^{\rm th}(z,H_{0})=\frac{1}{(1+z)H_0}.
\end{equation}

To carry out the age-redshift analysis of $\Lambda$CDM and $R_{\rm h}=ct$,
we will first attempt to fit the ages of 32 old passive galaxies distributed over the
redshift interval $0.117\leq z \leq 1.845$ (Simon et al. 2005), listed in Table~1 of
Samushia et al. (2010), assuming a uniform value of the time delay $\tau$ for
every galaxy. Following these authors, we will also assume a $12\%$
one-standard deviation uncertainty on the age measurements (Dantas et al. 2009,
2011; Samushia et al. 2010). The total sample is composed of three sub-samples:
10 field early-type galaxies from Treu et al. (1999, 2001, 2002), whose
ages were obtained by using the SPEED models of Nolan et al. (2001) and Jimenez et al. (2004);
20 red galaxies from the publicly released Gemini Deep Deep Survey (GDDS),
whose integrated light is fully dominated by evolved stars (Abraham et al.
2004; McCarthy et al. 2004); and the 2 radio galaxies LBDS 53W091 and
LBDS 53W069 (Dunlop et al. 1996; Spinrad et al. 1997). These data were
first collated by Jimenez et al. (2003).

\section{Optimization of the Model Parameters Using a Uniform $\tau$\label{sec:intro}}
In subsequent sections of this paper, we will study the impact of a
distributed incubation time on the overall fits to the data. However, in
this first simple approach (see, e.g., Dantas et al. 2009, 2011; Samushia et
al. 2010) we will assume an average delay factor $\langle\tau\rangle$ and
use it uniformally for every galaxy, so that we may compare our results
to those of previous work.

For each model, we optimize the fit by finding the set of parameters
($\mathbf{p}$) that minimize the $\chi^{2}$, using the statistic
\begin{eqnarray}
\chi^{2}_{age}(\tau, \mathbf{p})&=&\sum_{i=1}^{32}
\frac{\left[t^{\rm th}(z_{i},\mathbf{p})-t_{G}(z_{i})-\langle\tau\rangle\right]^{2}}{\sigma_{t_{G,i}}^{2}}\nonumber\\
&\equiv& A-2\ast\langle\tau\rangle \ast B+\langle\tau\rangle^{2}\ast C\;,
\end{eqnarray}
where $A\equiv \sum\left[t^{\rm th}(z_{i},\mathbf{p})-t_{G}(z_{i})\right]^{2}/\sigma_{t_{G,i}}^{2}$,
$B\equiv\sum\left[t^{\rm th}(z_{i},\mathbf{p})-t_{G}(z_{i})\right]/\sigma_{t_{G,i}}^{2}$, and
$C\equiv\sum1/\sigma_{t_{G,i}}^{2}$. The dispersions $\sigma_{t_{G,i}}$ represent the
uncertainties on the age measurements of the sample galaxies.  Given the form of Equation~(5),
we can marginalize $\langle\tau\rangle$ by minimizing $\chi^{2}_{age}$, which has a
minimum at $\langle\tau\rangle=B/C$, with a value $\hat{\chi}^{2}_{age}=A-B^{2}/C$. Note that
this procedure allows us to determine the optimized value of $\langle\tau\rangle$ along with the
best-fit parameters of the model being tested.

\subsection{$\Lambda$CDM}
In the concordance $\Lambda$CDM model, the dark-energy equation of state parameter,
$w_{\rm de}$, is exactly $-1$. The Universe is flat, $\Omega_{\rm de}=1-
\Omega_{\rm m}-\Omega_{\rm r}$, so there remain only two free parameters: $\Omega_{\rm m}$ and
$H_{0}$. Type Ia SN measurements (see, e.g., Garnavich et al. 1998;
Perlmutter et al.  1998, 1999; Riess et al. 1998; Schmidt et al. 1998),
CMB anisotropy data (see, e.g., Ratra et al. 1999; Podariu et al.
2001; Spergel et al. 2003; Komatsu et al. 2009, 2011; Hinshaw et al. 2013), and baryon
acoustic oscillation (BAO) peak length scale estimates (see, e.g.,
Percival et al. 2007; Gazta{\~n}aga et al.2009; Samushia \& Ratra 2009),
strongly suggest that we live in a spatially flat, dark energy-dominated universe
with concordance parameter values $\Omega_{\rm m}\approx0.27$ and
$H_{0}\approx70.0$ km $\rm s^{-1}$ $\rm Mpc^{-1}$.

In order to gauge how well $\Lambda$CDM and $R_{\rm h}=ct$ account for
the galaxy age-redshift measurements, we will first attempt to fit the data
with this concordance model, using prior values for all the parameters but one,
i.e., the unknown average delay time $\langle\tau\rangle$. We will improve the fit as much
as possible by marginalizing $\langle\tau\rangle$, as described above.  Fitting the 32 age-redshift
measurements with a theoretical $t^{\rm th}(z)$ function using
$\Omega_{\rm m}=0.27$ and $H_{0}=70.0$ km $\rm s^{-1}$ $\rm Mpc^{-1}$,
we obtain an optimized delay factor $\langle\tau\rangle=1.36$ Gyr, and a
$\chi^2_{\rm dof}=16.17/31=0.523$, remembering that all
of the $\Lambda$CDM parameters are assumed to have prior values, except for
$\langle\tau\rangle$.

\begin{figure}[h]
\centerline{\hskip0.9in\includegraphics[angle=0,scale=0.85]{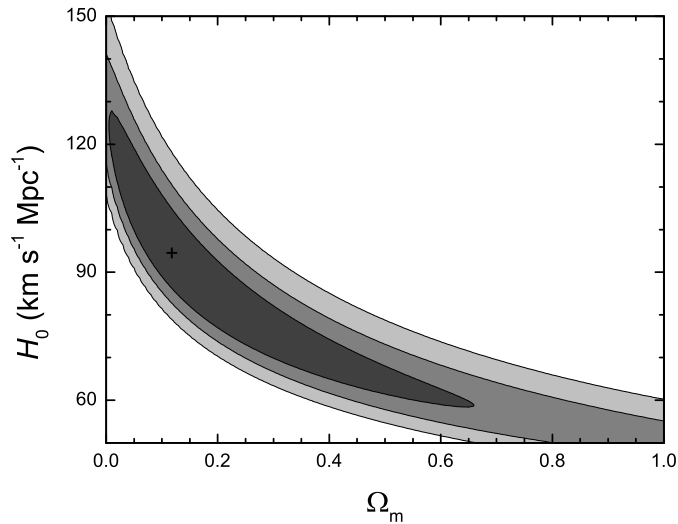}}
\caption{$1$-$3\sigma$-constraint contours for the flat $\Lambda$CDM
model, using the 32 age-redshift data. The cross indicates
the best-fit pair ($\Omega_{\rm m}$, $H_{0}$)=(0.12, 94.3).}\label{flat}
\end{figure}

If we relax the priors, and allow both $\Omega_{\rm m}$ and $H_{0}$
to be free parameters, we obtain best-fit values $(\Omega_{\rm m}$, $H_{0}$)
=(0.12, $94.3$ km $\rm s^{-1}$ $\rm Mpc^{-1}$), as illustrated in Figure~1. The
delay factor corresponding to this best fit is $\langle\tau\rangle=1.62$ Gyr, with a
$\chi_{\rm dof}^{2}=12.42/29=0.428$. Figure~1 also shows
the $1$-$3\sigma$ constraint contours of the probability function in the
$\Omega_{\rm m}$-$H_{0}$ plane. Insofar as the $\Lambda$CDM model is
concerned, this optimization has improved the quality of the fit as gauged
by the reduced  $\chi_{\rm dof}^{2}$, though the parameter values are
quite different from those of the concordance model. Nonetheless, these two
sets of values are still marginally consistent with each other because the data are
not good enough yet to improve the precision with which $\Omega_{\rm m}$
and $H_{0}$ are determined. It is also possible that treating
$\langle\tau\rangle$ as a uniform variable for all galaxies may be
over-constraining, but we will relax this condition in subsequent
sections and consider situations in which $\tau_i$ may be different
for each galaxy in the sample. We shall see that these optimized
values change quantitatively, though the qualitative results and
conclusions remain the same. The contours in Figure~1
show that at the $1\sigma$-level, we have $58.5<H_{0}<127.0$  km
$\rm s^{-1}$ $\rm Mpc^{-1}$, and $0.01<\Omega_{\rm m}<0.66$.
The cross indicates the best-fit pair. For the sake
of a direct one-on-one comparison between $\Lambda$CDM and
the $R_{\rm h}=ct$ Universe, the current status with these data
therefore suggests that we should use the concordance parameter
values, which are supported by many other kinds of measurements,
as described above.

\subsection{The $R_{\rm h}=ct$ Universe}
Regardless of what constituents may be present in the cosmic fluid,
insofar as the expansion dynamics is concerned, the $R_{\rm h}=ct$
Universe always has just one free parameter, $H_{0}$. The results of
fitting the age-redshift data with this cosmology are shown in Figure~2
(solid line). We see here that the best fit corresponds to $H_{0}=
67.2_{-4.0}^{+4.5}$ km $\rm s^{-1}$ $\rm Mpc^{-1}$ ($1\sigma$).
The corresponding delay factor is $\langle\tau\rangle=2.72$ Gyr.  With $32-2=30$
degrees of freedom, we have $\chi_{\rm dof}^{2}=13.05/30=0.435$.

To facilitate a direct comparison between $\Lambda$CDM and $R_{\rm h}=ct$,
we show in Figure~3 the galaxy ages (i.e., $t_{G}+\langle\tau\rangle$),
together with the best-fit theoretical curves for the $R_{\rm h}=ct$
Universe (with $H_{0}=67.2$ km $\rm s^{-1}$ $\rm Mpc^{-1}$
and $\langle\tau\rangle=2.72$ Gyr), the concordance model (with
$\langle\tau\rangle=1.36$ Gyr and prior values for all the other
parameters), and for the optimized $\Lambda$CDM model (with
$H_{0}=94.3$ km $\rm s^{-1}$ $\rm Mpc^{-1}$, $\Omega_{\rm m}
=0.12$, and $\langle\tau\rangle=1.62$ Gyr). As described above,
$\langle\tau\rangle$ is the average incubation time or delay factor,
which accounts for our ignorance concerning the amount of time
elapsed since the big bang to the initial formation of the object.
At the very minimum, $\langle\tau\rangle$ must be greater than $\sim 300$ Myr,
this being the time at which Population III stars would have
established the necessary conditions for the subsequent formation
of Population II stars (see, e.g., Melia 2014b and references cited therein).
The galaxies could not have formed any earlier than this, based on the
physics we know today.

\begin{figure}[h]
\centerline{\includegraphics[angle=0,scale=0.8]{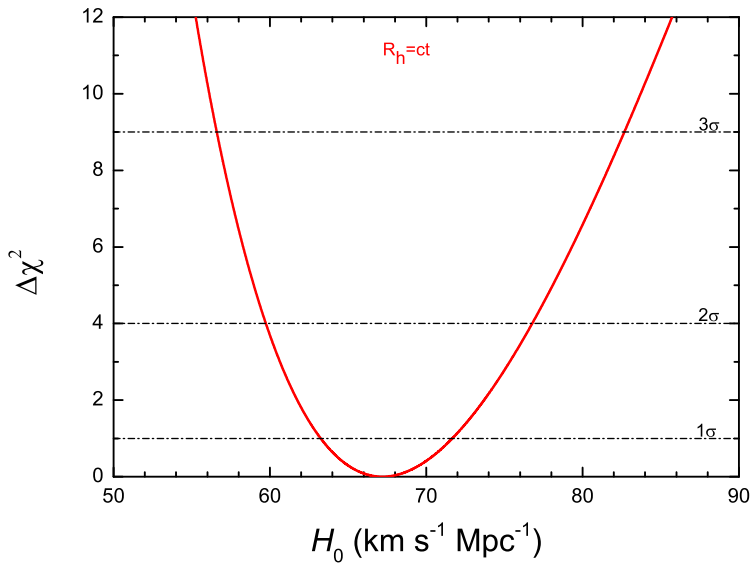}}
\caption{Constraints on the Hubble constant, $H_{0}$, in the
context of $R_{\rm h}=ct$.}\label{Rh}
\end{figure}

\begin{figure}[hp]
\centerline{\includegraphics[angle=0,scale=0.8]{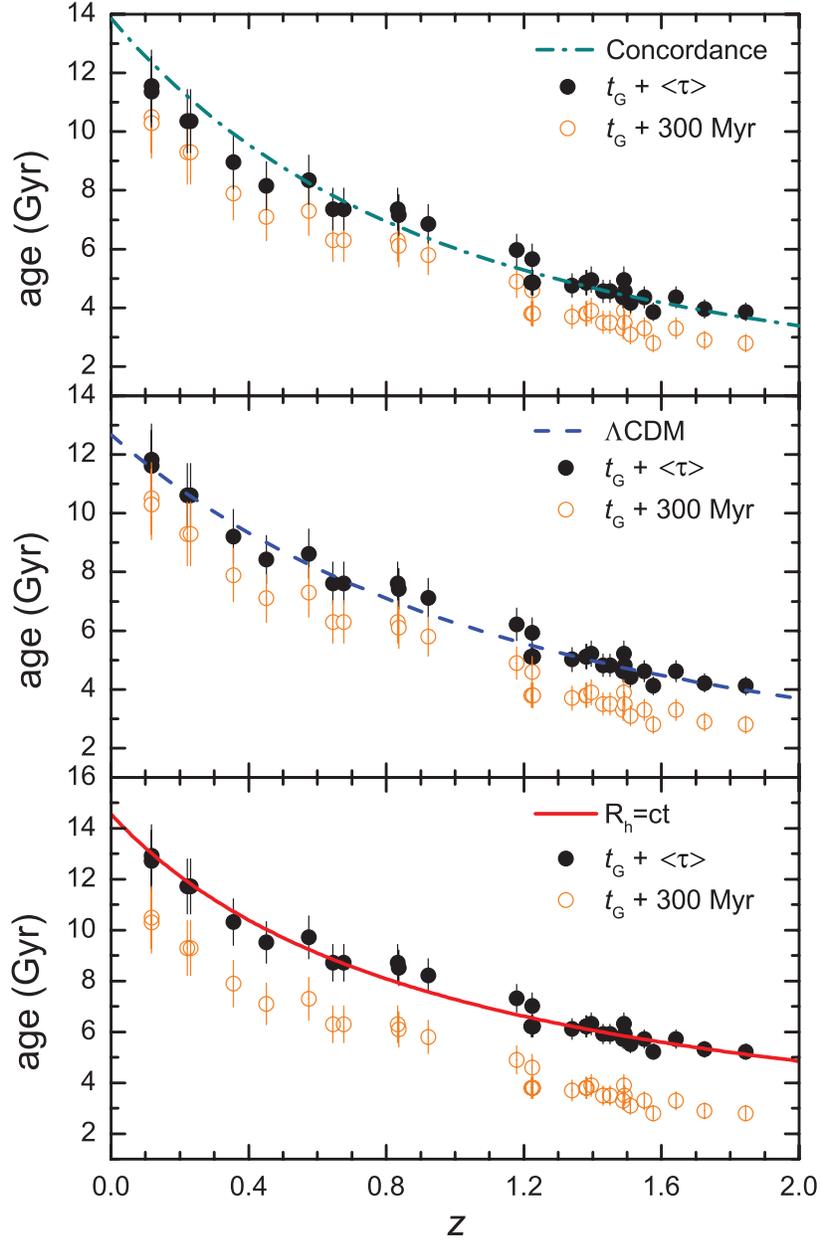}}
\caption{The complete age-redshift sample (\emph{solid points}), and the
best-fit theoretical curves: (\emph{dot-dashed line}) the concordance model,
with its sole optimized parameter $\langle\tau\rangle=1.36$ Gyr; (\emph{dashed line}) the
standard, flat $\Lambda$CDM cosmology, with optimized parameters
$\Omega_{\rm m}=0.12$, $H_{0}=94.3_{-35.8}^{+32.7}$ ($1\sigma$) km
$\rm s^{-1}$ $\rm Mpc^{-1}$, and $\langle\tau\rangle=1.62$ Gyr; (\emph{solid line}) the
$R_{\rm h}=ct$ Universe, with $H_{0}=67.2_{-4.0}^{+4.5}$
($1\sigma$) km $\rm s^{-1}$ $\rm Mpc^{-1}$ and $\langle\tau\rangle=2.72$ Gyr.
The \emph{empty circles} show the minimum ages the galaxies could
have using $t_G(z_{i})+300$ Myr.}\label{age}
\end{figure}

In this figure, we also show $t_{G}+300$ Myr versus $z$ (circles)
to illustrate the minimum possible ages the galaxies could have, given
what we now know about the formation of Population II and Population
III stars. We note that the best fit value of $\langle\tau\rangle$, in both
$\Lambda$CDM and $R_{\rm h }=ct$, is fully consistent with the
supposition that all of the galaxies should have formed after the
transition from Population III to Population II
stars at $t\sim300$ Myr. Strictly based on their $\chi^{2}_{\rm dof}$
values, the concordance $\Lambda$CDM model and the
$R_{\rm  h}=ct$ Universe appear to fit the passive galaxy
age-redshift relationship (i.e., $t_{G}+\langle\tau\rangle$ versus $z$)
comparably well. However, because these models formulate their observables
(such as the theoretical ages in Equations~1 and 4) differently, and
because they do not have the same number of free parameters,
a comparison of the likelihoods for either being closer to the `true' model
must be based on model selection tools, which we discuss in \S~5
below. But first, we will strengthen this analysis by considering
possibly more realistic distributions of the delay time $\tau$.

\section{Optimization of the Model Parameters Using a Distributed $\tau$}
We now relax the constraint that the time delay should have the same value
$\langle\tau\rangle$ for every galaxy, and instead consider two
representative distributions: (i) a Gaussian
\begin{equation}
P(\tau)\propto \exp{-{(\tau-\tau_c)^2\over 2\sigma_\tau^2}}\;,
\end{equation}
where the mean value $\tau_c$ is to be optimized for each theoretical
fit, given some dispersion $\sigma_\tau$; and (ii) a top-hat
\begin{equation}
P(\tau)=\begin{cases}
                  {\rm const} & (\tau_c-\sigma_\tau<\tau<\tau_c+\sigma_\tau)\\
                  0 & ({\rm otherwise})\;,
              \end{cases}
\end{equation}
with $\sigma_\tau$ now representing the width of the distribution.

\begin{figure}[hp]
\centerline{\includegraphics[angle=0,scale=1.2]{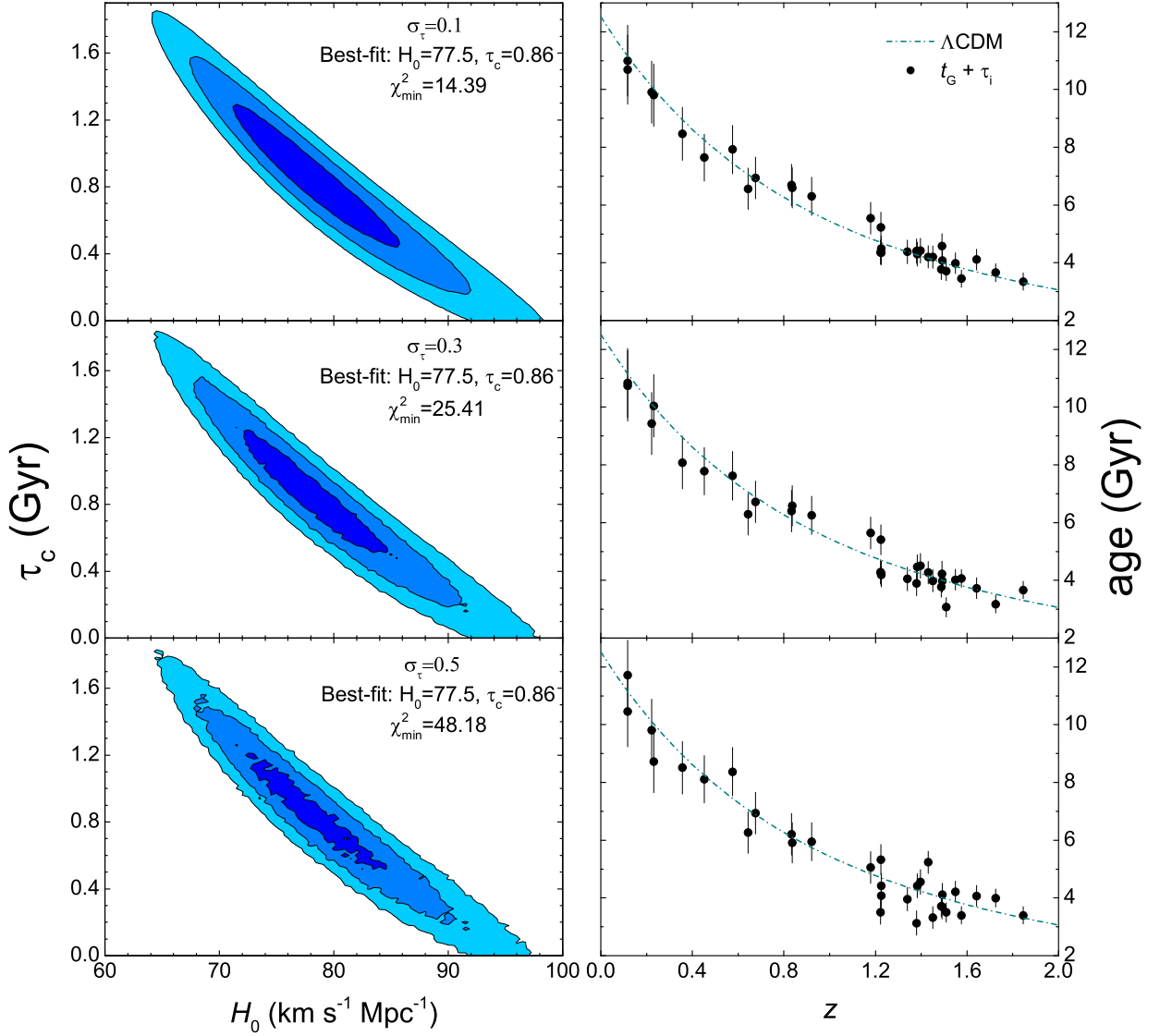}}
\caption{$\Lambda$CDM with a Gaussian distribution of $\tau$ values.
Left-hand panels: fitted values of $\tau_c$ and $H_0$ using the
complete age-redshift sample (right-hand panels; \emph{solid points}).
The theoretical curves (right-hand panel; \emph{dot-dashed curves}) correspond
to the parameter values that minimize the $\chi^2$ (shown on the left).
For $\Lambda$CDM, a Gaussian distribution in $\tau$ results in an
optimized value of the Hubble constant ($\sim 77.5$  km $\rm s^{-1}$
$\rm Mpc^{-1}$) only weakly dependent on $\sigma_\tau$, and a
mean delay time $\tau_c\sim 0.86$ Gyr.}
\end{figure}

To isolate the various influences as much as possible, we begin with the
concordance $\Lambda$CDM model, for which $\Omega_{\rm m}=0.27$ and
$w_{\rm de}=-1$ (i.e., dark energy is assumed to be a cosmological constant),
though we optimize the Hubble constant to maximize the quality of the fit. For each
distribution $P(\tau)$ and assumed value of $\sigma_\tau$, we randomnly
assign the time delay $\tau_i$ to each galaxy and then find the best-fit
values of $H_0$ and $\tau_c$ by minimizing the $\chi^2$ using the
statistic
\begin{equation}
\chi^{2}_{age}(\tau_c, \mathbf{p})=\sum_{i=1}^{32}
\frac{\left[t^{\rm th}(z_{i},\mathbf{p})-t_{G}(z_{i})-\tau_i(\tau_c,\sigma_\tau)
\right]^{2}}{\sigma_{t_{G,i}}^{2}}\;.
\end{equation}
The left-hand panels of Figure~4 show
the ensuing distributions of $\tau_c$ and $H_0$ values for a Gaussian $P(\tau)$,
and three different assumed dispersions $\sigma_\tau$. In this case, the
optimized Hubble constant ($\sim 77.5$  km $\rm s^{-1}$ $\rm Mpc^{-1}$)
is effectively independent of $\sigma_\tau$, while the best-fit value of the
mean delay time $\tau_c$ is restricted to $\sim0.86$
Gyr. Not surprisingly, the scatter about the best-fit theoretical curve
worsens as $\sigma_\tau$ increases, resulting in larger values of $\chi_{\rm min}^2$.

\begin{figure}[hp]
\centerline{\includegraphics[angle=0,scale=1.2]{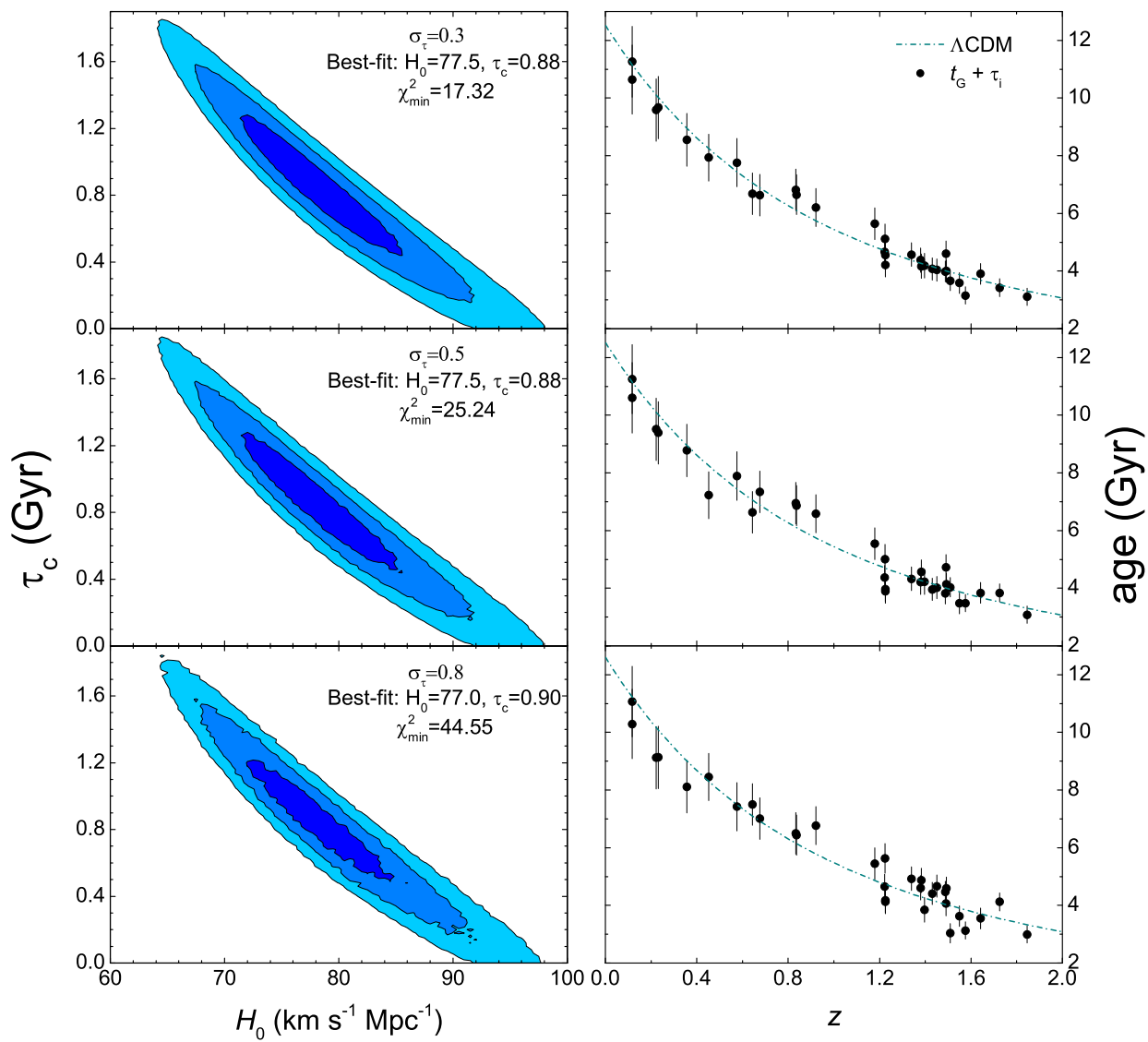}}
\caption{Same as Figure~4, except now for the top-hat
distribution $P(\tau)$ given in Equation~(7). In this case, both
$H_0$ and $\tau_c$ are effectively independent of the assumed
distribution width $\sigma_\tau$.}
\end{figure}

Assuming a top-hat $P(\tau)$ with $\Lambda$CDM produces the results
shown in Figure~5. The best-fit values of $H_0$ and $\tau_c$ are very
similar to those associated with Figure~4. In this case, both $H_0$ and $\tau_c$
are effectively independent of the assumed distribution width $\sigma_\tau$.

We follow the same procedure for the $R_{\rm h}=ct$ Universe, first
considering a Gaussian distribution of $\tau_i$ values (Figure~6), followed by the
top-hat distribution (Figure~7). The comparison between these two cosmological
models may be summarized as follows: the best-fit results are very
similar for both the Gaussian and top-hat $P(\tau)$ distributions, for the same cosmological
model; strictly based on their minimum $\chi^{2}$ values, the concordance $\Lambda$CDM
model and the $R_{\rm  h}=ct$ Universe appear to fit the passive galaxy
age-redshift relationship (i.e., $t_{G}+\tau_{i}$ versus $z$) comparably well,
independent of what kind of time-delay distribution is assumed.

\begin{figure}[hp]
\centerline{\includegraphics[angle=0,scale=1.2]{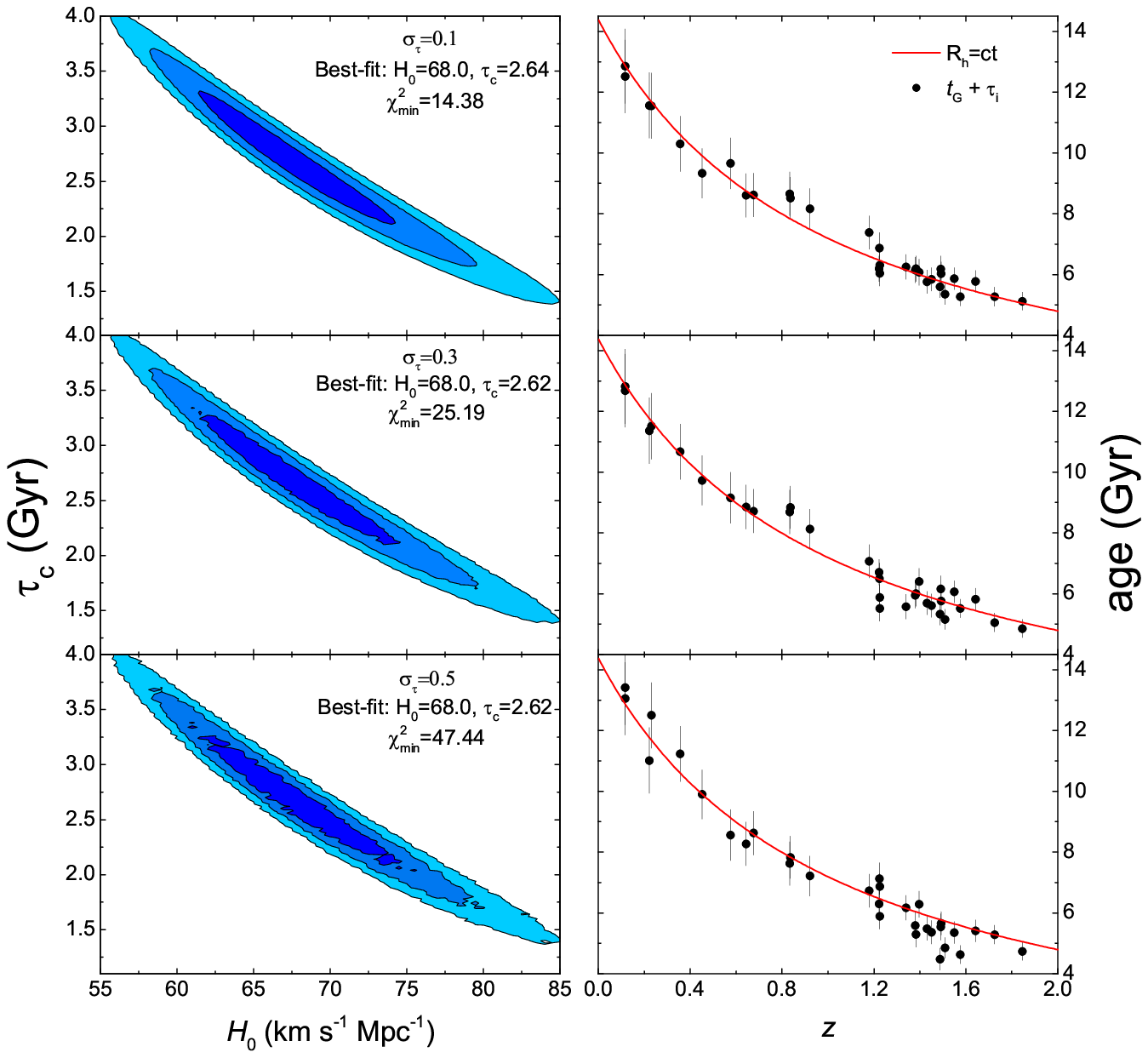}}
\caption{Same as Figure~4, except now for the $R_{\rm h}=ct$
Universe.}
\end{figure}

\begin{figure}[hp]
\centerline{\includegraphics[angle=0,scale=1.2]{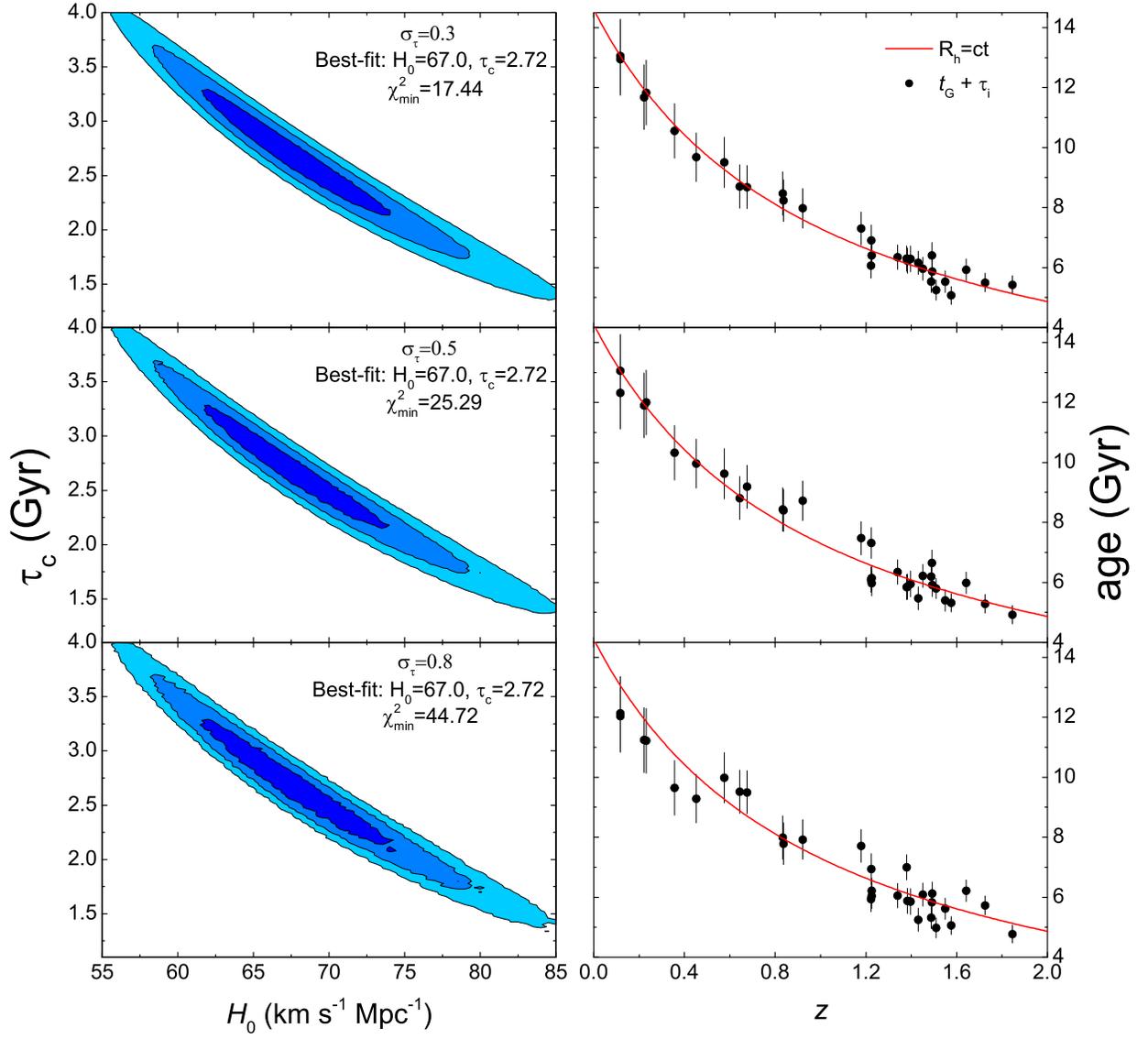}}
\caption{Same as Figure~6, except now for the top-hat
distribution $P(\tau)$ given in Equation~(7).}
\end{figure}

\begin{deluxetable}{lccc}
\tablewidth{210pt}
\tabletypesize{\footnotesize}
\tablecaption{AIC Likelihood Estimation}\tablenum{1}
\tablehead{Distributed $\tau$&\colhead{$\sigma_{\tau}$}&\colhead{$\Lambda$CDM}&\colhead{$R_{\rm h}=ct$} }
\startdata
Gaussian & 0.1 & $50\%$ & $50\%$ \\
         & 0.3 & $47\%$ & $53\%$ \\
         & 0.5 & $41\%$ & $59\%$ \\
Top-hat  & 0.3 & $51\%$ & $49\%$ \\
         & 0.5 & $50\%$ & $50\%$ \\
         & 0.8 & $52\%$ & $48\%$ \\
\enddata
\end{deluxetable}

\section{Model Selection Tools}
Several model selection tools used in cosmology (see,
e.g., Melia \& Maier 2013, and references cited therein) include
the Akaike Information Criterion, ${\rm AIC}=\chi^{2}+2n$, where
$n$ is the number of free parameters (Liddle 2007),
the Kullback Information Criterion, ${\rm KIC}=\chi^{2}+3n$ (Cavanaugh
2004), and the Bayes Information Criterion,
${\rm BIC}=\chi^{2}+(\ln N)n$, where $N$ is the number of data points
(Schwarz 1978). In the case of AIC, with ${\rm AIC}_\alpha$
characterizing model $\mathcal{M}_\alpha$,
the unnormalized confidence that this model is true is the Akaike
weight $\exp(-{\rm AIC}_\alpha/2)$. Model $\mathcal{M}_\alpha$ has likelihood
\begin{equation}
P(\mathcal{M}_\alpha)= \frac{\exp(-{\rm AIC}_\alpha/2)}
{\exp(-{\rm AIC}_1/2)+\exp(-{\rm AIC}_2/2)}
\end{equation}
of being the correct choice in this one-on-one comparison. Thus, the difference
$\Delta \rm AIC \equiv {\rm AIC}_2\nobreak-{\rm AIC}_1$ determines the extent to which $\mathcal{M}_1$
is favoured over~$\mathcal{M}_2$. For Kullback
and Bayes, the likelihoods are defined analogously.

For the case of the average delay factor $\langle\tau\rangle$,
with the optimized fits we have reported in this paper, our analysis of the
age-\emph{z} shows that the KIC does not favour either $R_{\rm h}=ct$
or the concordance model when we assume prior values for all of its parameters.
The calculated KIC likelihoods in this case are $\approx 51.5\%$ for $R_{\rm h}=ct$,
versus $\approx 48.5\%$ for $\Lambda$CDM. However, if we relax some of the priors,
and allow both $\Omega_{\rm m}$ and $H_0$ to be optimized in $\Lambda$CDM, then
$R_{\rm h}=ct$ is favoured over the standard model with a likelihood of
$\approx 66.5\%$ versus $33.5\%$ using AIC, $\approx 76.6\%$ versus $\approx 23.4\%$
using KIC, and $\approx 80.5\%$ versus $\approx 19.5\%$ using BIC.

For the distributed time delays discussed in Section~4, the model selection
criteria result in the likelihoods shown in Table~1. Note that in this case, both
the concordance $\Lambda$CDM and $R_{\rm h}=ct$ models have the same
free parameters (i.e., $H_{0}$ and $\tau_{c}$), so the information criteria
should all provide the same results. For the sake of clarity, we therefore
show only the AIC results in Table~1, where we see that the AIC does
not favour either $R_{\rm h}=ct$ or the concordance model, regardless
of which distribution is adopted for the incubation time.

\section{Numerical Simulations\label{sec:disc}}
The results of our analysis suggest that the measurement of galaxy ages may
be used to identify the preferred model in a one-on-one comparison.
In using the model selection tools, the outcome $\Delta\equiv$ AIC$_1-$
AIC$_2$ (and analogously for KIC and BIC) is judged `positive' in the range
$\Delta=2-6$, and `strong' for $\Delta>6$. As we have seen, the
adoption of prior values for the parameters in $\Lambda$CDM produces
comparable likelihood outcomes for both models, regardless of whether
we assume a uniform time delay $\langle\tau\rangle$, or a distribution
of values. Of course, a proper statistical comparison between $\Lambda$CDM
and $R_{\rm h}=ct$ should not have to rely on prior values, particularly
since the optimized cosmological parameters differ from survey to survey.
If we don't assume prior values for the parameters in $\Lambda$CDM,
and optimize them to produce a best fit to the galaxy-age data, the
corresponding $\Delta$ using the currently known 32 galaxy ages
falls within the `positive' range in favor of $R_{\rm h}=ct$,
though not yet the strong one. These
results are therefore suggestive, but still not sufficient to rule out
either model. In this section, we will therefore estimate the
sample size required to significantly strengthen the
evidence in favour of $R_{\rm h}=ct$ or $\Lambda$CDM, by conservatively
seeking an outcome even beyond $\Delta\simeq11.62$, i.e., we will see what is
required to produce a likelihood $\sim 99.7\%$ versus $\sim 0.3\%$,
corresponding to $3\sigma$.

Since the results do not appear to depend strongly on whether one
chooses a uniform $\langle\tau\rangle$ for all the galaxies, or a distribution
of individual $\tau_i$ values, we will first use the same average value
$\langle\tau\rangle$ in our simulations, and then discuss how these results
would change for a distributed incubation time.
We will consider two cases: one in which the background cosmology is
assumed to be $\Lambda$CDM, and a second in which it is $R_{\rm h}=ct$,
and we will attempt to estimate the number of galaxy ages
required in each case in order to rule out the alternative (incorrect)
model at a $\sim 99.7\%$ confidence level. The synthetic galaxy ages
are each characterized by a set of parameters denoted as ($z$, $t[z]$),
where $t(z)=t_{G}+\langle\tau\rangle$. We generate the synthetic sample
using the following procedure:

1. Since the current 32 old passively evolving galaxies are distributed over
the redshift interval $0.117 \leq z \leq 1.845$, we assign $z$ uniformly
between $0.1$ and $2.0$.

2. With the mock $z$, we first infer $t(z)$ from Equations~(1) and (4)
corresponding either to a flat $\Lambda$CDM cosmology with $\Omega_{\rm m}=
0.27$ and $H_{0}=70$ km $\rm s^{-1}$ $\rm Mpc^{-1}$ (\S~4.2), or the
$R_{\rm h}=ct$ Universe with $H_{0}=70$ km $\rm s^{-1}$ $\rm Mpc^{-1}$
(\S~4.1). We then assign a deviation ($\Delta t$) to the $t(z)$ value, i.e.,
we infer $t'(z)$ from a normal distribution whose center value is $t(z)$ and
$\sigma=0.35$ is its deviation (see Bengaly et al. 2014). The typical value
of $\sigma=0.35$ is taken from the current (observed) sample, which yields
a mean and median deviation of $\sigma=0.38$ and $0.33$, respectively.

3. Since the observed error $\sigma_{t}$ is about $12\%$ of the age measurement,
we will also assign a dispersion $\sigma_{t}=0.12\;t'(z)$ to the synthetic sample.

This sequence of steps is repeated for each galaxy in the sample, which
is enlarged until the likelihood criterion discussed above is reached. As with
the real 32-ages sample, we optimize the model fits by minimizing the $\chi^{2}$ function
$\chi^{2}=\sum\left\{\left[t^{\rm th}(z_{i},\mathbf{p})-t'(z_{i})\right]^{2}/\sigma_{t,i}^{2}\right\}$.
This minimization is equivalent to maximizing the likelihood function $\mathcal{L}\propto
\rm exp\left(-\chi^{2}/2\right)$. We employ Markov-chain Monte Carlo techniques.
In each Markov chain, we generate $10^5$ samples according to the likelihood function.
Then we derive the cosmological parameters from a statistical analysis of the sample.

\begin{figure}[hp]
\centerline{\includegraphics[angle=0,scale=0.8]{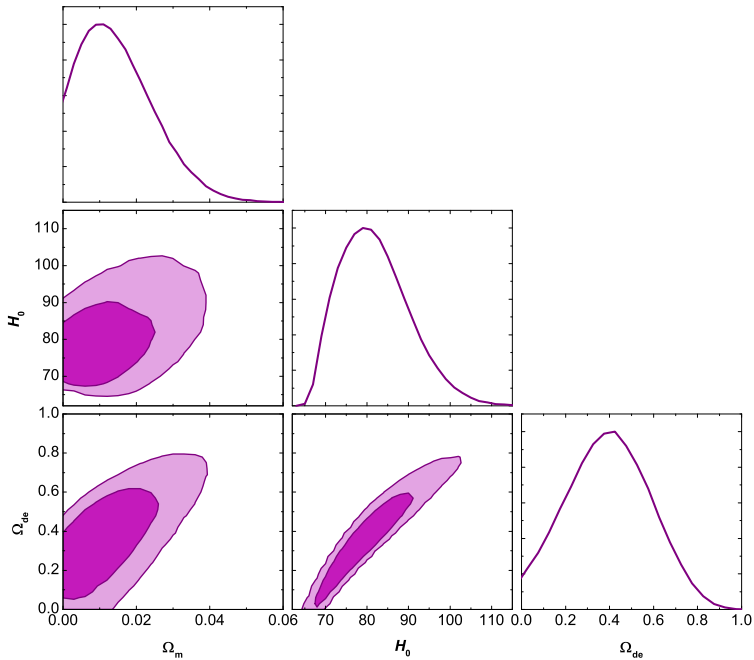}}
\vskip-0.2in
\caption{The 1-D probability distributions and 2-D regions with the $1\sigma$ and
$2\sigma$ contours corresponding to the parameters $\Omega_{\rm m}$, $\Omega_{\rm de}$,
and $H_{0}$ in the best-fit $\Lambda$CDM model, using the simulated sample with
350 ages, assuming $R_{\rm h}=ct$ as the background cosmology.}
\end{figure}

\subsection{Assuming $R_{\rm h}=ct$ as the Background Cosmology}
We have found that a sample of at least 350 galaxy ages is required
in order to rule out $\Lambda$CDM at the $\sim 99.7 \%$ confidence level. The
optimized parameters corresponding to the best-fit $\Lambda$CDM model for
these simulated data are displayed in Figure~8. To allow for the greatest
flexibility in this fit, we relax the assumption of flatness, and allow
$\Omega_{\rm de}$ to be a free parameter, along with $\Omega_{\rm m}$. Figure~8
shows the 1-D probability distribution for each parameter ($\Omega_{\rm m}$,
$\Omega_{\rm de}$, $H_{0}$), and 2-D plots of the $1\sigma$ and
$2\sigma$ confidence regions for two-parameter combinations. The best-fit
values for $\Lambda$CDM using the simulated sample with 350 ages
in the $R_{\rm h}=ct$ Universe are $\Omega_{\rm m}=0.011$,
$\Omega_{\rm de}=0.37_{-0.32}^{+0.25}$ $(1\sigma)$, and $H_{0}=79.9_{-12.7}^{+10.7}$
$(1\sigma)$ km $\rm s^{-1}$ $\rm Mpc^{-1}$. Note that the simulated ages provide
a good constraint on $\Omega_{\rm de}$, but only a weak one on $\Omega_{\rm m}$;
only an upper limit of $\sim0.025$ can be set at
the $1\sigma$ confidence level.

\begin{figure}[hp]
\centerline{\hskip 0.5in\includegraphics[angle=0,scale=0.7]{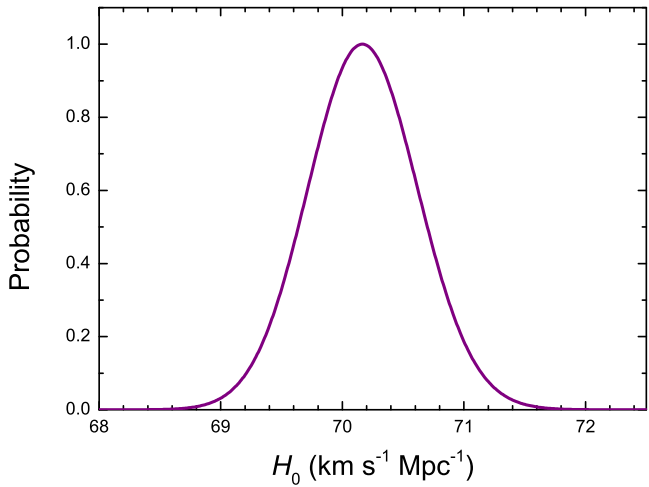}}
\vskip-0.2in
\caption{The 1-D probability distribution for the parameter $H_{0}$ in the $R_{\rm h}=ct$ universe,
using a sample of 350 ages, simulated with $R_{\rm h}=ct$ as the
background cosmology. The assumed value for $H_0$ in the simulation was
$H_{0}=70$ km $\rm s^{-1}$ $\rm Mpc^{-1}$.}
\end{figure}

In Figure~9, we show the corresponding 1-D probability distribution of $H_{0}$ for
the $R_{\rm h}=ct$ universe. The best-fit value for the simulated sample is
$H_{0}=70.2_{-0.5}^{+0.5}$ $(1\sigma)$ km $\rm s^{-1}$ $\rm Mpc^{-1}$. The
assumed value for $H_0$ in the simulation was $H_{0}=70$ km $\rm s^{-1}$ $\rm Mpc^{-1}$.

Since the number $N$ of data points in the sample is now much greater than one, the
most appropriate information criterion to use is the BIC. The logarithmic penalty
in this model selection tool strongly suppresses overfitting if $N$ is large
(the situation we have here, which is deep in the asymptotic regime). With $N=350$,
our analysis of the simulated sample shows that the BIC would favour the $R_{\rm h}=ct$
Universe over $\Lambda$CDM by an overwhelming likelihood of $99.7\%$ versus only $0.3\%$
(i.e., the prescribed $3\sigma$ confidence limit).

\subsection{Assuming $\Lambda$CDM as the Background Cosmology}
In this case, we assume that the background cosmology is $\Lambda$CDM,
and seek the minimum sample size to rule out $R_{\rm h}=ct$ at the
$3\sigma$ confidence level. We have found that a minimum of 45 galaxy
ages are required to achieve this goal. To allow for the greatest flexibility
in the $\Lambda$CDM fit, here too we relax the assumption of flatness, and
allow $\Omega_{\rm de}$ to be a free parameter, along with $\Omega_{\rm m}$.
In Figure~10, we show the 1-D probability distribution for each parameter
($\Omega_{\rm m}$, $\Omega_{\rm de}$, $H_{0}$), and 2-D plots of
the $1\sigma$ and $2\sigma$ confidence regions for two-parameter
combinations. The best-fit values for $\Lambda$CDM using this simulated sample
with 45 galaxy ages are $\Omega_{\rm m}=0.28_{-0.11}^{+0.12}$ $(1\sigma)$,
$\Omega_{\rm de}=0.30$, and $H_{0}=60.1_{-7.6}^{+9.2}$ $(1\sigma)$ km
$\rm s^{-1}$ $\rm Mpc^{-1}$. Note that the simulated ages now give a good
constraint on $\Omega_{\rm m}$, but only a weak one on $\Omega_{\rm de}$;
only an upper limit of $\sim0.83$ can be set at the $1\sigma$ confidence level.

The corresponding 1-D probability distribution of $H_{0}$ for the $R_{\rm h}=ct$
universe is shown in Figure~11. The best-fit value for the simulated sample is
$H_{0}=85.0_{-1.5}^{+1.6}$ $(1\sigma)$ km $\rm s^{-1}$ $\rm Mpc^{-1}$. This is
similar to that in the standard model, but not exactly the same, reaffirming
the importance of reducing the data separately for each model being tested.
With $N=45$, our analysis of
the simulated sample shows that in this case the BIC would favour $\Lambda$CDM
over $R_{\rm h}=ct$ by an overwhelming likelihood of $99.7\%$ versus only $0.3\%$
(i.e., the prescribed $3\sigma$ confidence limit).

\begin{figure}[hp]
\centerline{\includegraphics[angle=0,scale=0.8]{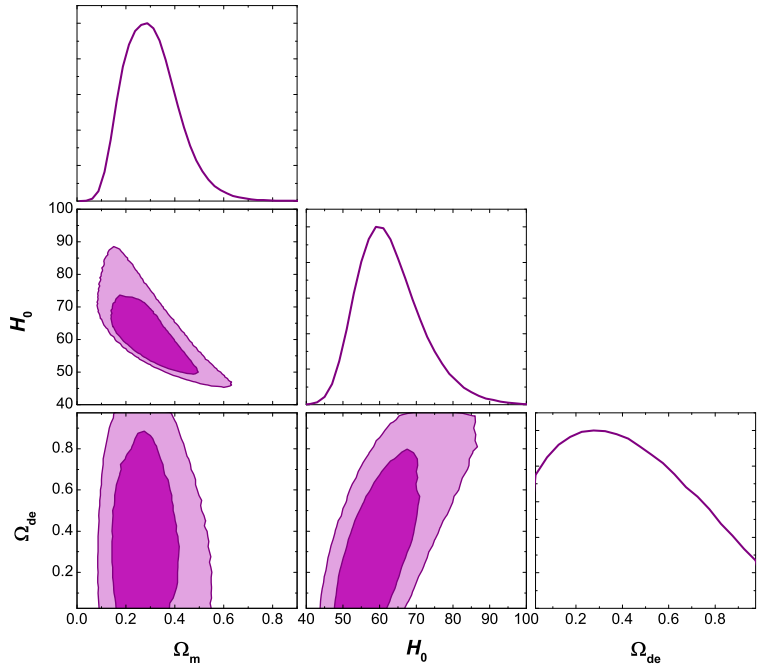}}
\vskip-0.2in
\caption{Same as Figure~8, except now with a flat $\Lambda$CDM as the (assumed)
background cosmology. The simulated model parameters were $\Omega_{\rm m}=0.27$
and $H_{0}=70$ km $\rm s^{-1}$ $\rm Mpc^{-1}$.}
\end{figure}

These results were obtained assuming a uniform incubation time $\langle\tau
\rangle$ throughout the mock sample. Of course, if the incubation time is
distributed, the corresponding uncertainty in its distribution function
will contribute to the variance of the ``observed" age of the Universe.
Thus, adding the scatter in the uncertain incubation time $\tau_{c}$ to
the simulations would change the constructed sample size required to achieve
the $3\sigma$ results discussed above. We have therefore carried out
additional simulations using a Gaussian $P(\tau)$ distribution of the incubation
time, and an assumed dispersion $\sigma_{\tau}=0.3$. The corresponding
uncertainty in $\tau$ contributes to the variance of the ``observed" age
of the Universe. From these results, we estimate that a sample of about
55 galaxy ages would be needed to rule out $R_{\rm h}=ct$ at a $\sim 99.7\%$
confidence level if the real cosmology were $\Lambda$CDM, while a sample
of at least 500 ages would be needed to similarly rule out $\Lambda$CDM
if the background cosmology were instead $R_{\rm h}=ct$.

\section{Discussion and Conclusions\label{sec:disc}}
In this paper, we have used the sample of high-redshift galaxies with
measured ages to compare the predictions of several cosmological models. We
have individually optimized the parameters in each case by minimizing the $\chi^{2}$
statistic. Using a sample of 32 passively evolving galaxies distributed over the
redshift interval $0.117\leq z\leq1.845$, we have demonstrated how these
age-redshift data can constrain parameters, such as $H_{0}$ and $\Omega_{\rm m}$.
For $\Lambda$CDM, these data are not good enough to improve upon the concordance
values yet, but are approaching the probative levels seen with currently available
gamma-ray burst luminosity data (Wei et al. 2013), strong gravitational-lensing
measurements (see, e.g., Suyu et al. 2013), and measurements of the
Hubble parameter as a function of redshift (Melia \& Maier 2013).

\begin{figure}[hp]
\centerline{\hskip 0.5in\includegraphics[angle=0,scale=0.7]{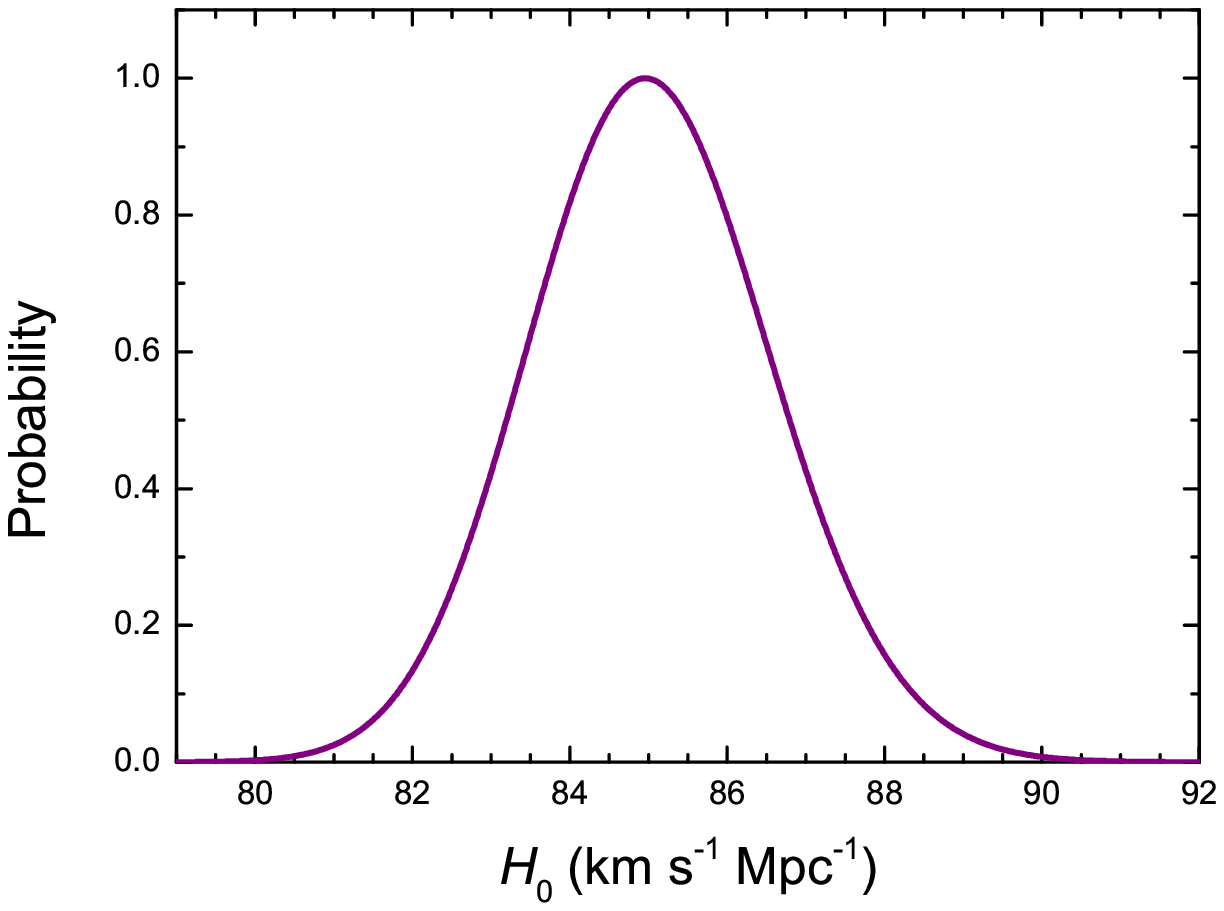}}
\vskip-0.2in
\caption{Same as Figure~9, except now with $\Lambda$CDM as the (assumed)
background cosmology.}
\end{figure}

Based solely on these 32 passively evolving galaxies, a comparison of the
$\chi^2_{\rm dof}$ for the $R_{\rm h}=ct$ Universe and the concordance
$\Lambda$CDM model shows that the age-redshift data do not yet favour either
model. The $R_{\rm h}=ct$ Universe fits the data with $\chi^2_{\rm dof}=0.435$ for
a Hubble constant $H_{0}=67.2_{-4.0}^{+4.5}$ km $\rm s^{-1}$ $\rm Mpc^{-1}$
and a average delay time $\langle\tau\rangle=2.72$ Gyr. By comparison, the concordance model
fits these same data with a reduced $\chi^2_{\rm dof}=0.523$, with
a delay time $\langle\tau\rangle=1.36$ Gyr. Both are consistent with the view
that none of these galaxies should have started forming prior to the
transition from Population III to Population II stars at $\sim 300$ Myr.
However, if we relax some of the priors,
and allow both $\Omega_{\rm m}$ and $H_0$ to be optimized in $\Lambda$CDM, we
obtain best-fit values $\Omega_{\rm m}=0.12_{-0.11}^{+0.54}$ and
$H_{0}=94.3_{-35.8}^{+32.7}$ km $\rm s^{-1}$ $\rm Mpc^{-1}$. The delay factor
corresponding to this best fit is $\langle\tau\rangle=1.62$ Gyr, with a $\chi^2_{\rm dof}=0.428$.
The current sample favours $R_{\rm h}=ct$ over the standard model
with a likelihood of $\approx 66.5\%-80.5\%$ versus $\approx 19.5\%-33.5\%$.

We also analyzed the age-redshift relationship in cases where the
delay factor $\tau$ may be different from galaxy to galaxy, and considered
two representative distributions: (i) a Gaussian; and (ii) a top-hat. We
found that the optimized cosmological parameters change quantitatively,
though the qualitative results and conclusions remain the same, independent of
what kind of the distribution one assumes for $\tau$. Though one does
not in reality expect the delay factor to be uniform, the fact that its
distribution does not significantly affect the results can be useful
for a qualitative assessment of the data. It also suggests that the outcome
of our analysis is insensitive to the underlying assumptions we have made.

But though galaxy age estimates currently tend to slightly favour $R_{\rm h}=ct$
over $\Lambda$CDM, the known sample of such measurements is still too small
for us to completely rule out either model. We have therefore considered
two synthetic samples with characteristics similar to those of the 32 known
age measurements, one based on a $\Lambda$CDM background cosmology, the other on
$R_{\rm h}=ct$. From the analysis of these simulated ages, we have
estimated that a sample of about $45-55$ galaxy ages would be needed to rule out
$R_{\rm h}=ct$ at a $\sim 99.7\%$ confidence level if the real cosmology
were in fact $\Lambda$CDM, while a sample of $350-500$ ages would
be needed to similarly rule out $\Lambda$CDM if the background cosmology
were instead $R_{\rm h}=ct$. These ranges allow for the possible
contribution of an uncertainty in $\tau$ to the variance of the observed
age of the Universe at each redshift. The difference in required sample size
is due to $\Lambda$CDM's greater flexibility in fitting the data, since
it has a larger number of free parameters.

Both the Gaussian and Top-hat distributions that we have incorporated
into this study have assumed that the mean and scatter of the incubation
time are constant with redshift. However, it would not be unreasonable
to suppose that these quantities could have a systematic dependence on $z$.
To examine how the results might change in this case, we have
therefore also analyzed the real data using a Gaussian distribution
\begin{equation}
P(\tau)\propto \exp{-{\left[\tau-\tau_c\cdot(1+z_{i})^{\alpha}\right]^2\over
2\left[\sigma_\tau\cdot(1+z_{i})^{\alpha}\right]^2}}\;,
\end{equation}
with $\sigma_\tau=0.3$ and, for simplicity, $\alpha=1$.
Figure~12(a) shows the corresponding distributions of $\tau_{c}$ and
$H_{0}$ for the concordance $\Lambda$CDM model, with best-fit
values $(H_{0}, \tau_{c})=(82.5, 0.26)$. The analogous distributions
for the $R_{\rm h}=ct$ Universe are shown in Figure~12(b). In this
case, the best fit corresponds to $(H_{0}, \tau_{c})=(80.0, 0.73)$.
The added redshift dependence has not changed the result that both
models fit the passive galaxy age-redshift relationship comparably
well, based solely on their reduced $\chi^2$'s. Note that in this case,
both the concordance $\Lambda$CDM and $R_{\rm h}=ct$ models have the
same free parameters (i.e., $(H_{0}$ and $\tau_{c}$), so the information
criteria should all provide the same results. Therefore, we show only
the AIC results here. We find that the AIC does not favour either
$R_{\rm h}=ct$ or the concordance $\Lambda$CDM model, with relatively likelihoods
of $\approx 54\%$ versus $\approx 46\%$.
Note, however, that the best-fit value of $\tau_{c}$ in $\Lambda$CDM
does not appear to be consistent with the supposition that all of
the galaxies should have formed after the transition from Population~III
to Population~II stars at $t\sim300$ Myr.

\begin{figure*}
\hskip -0.04in
\includegraphics[angle=0,scale=0.7]{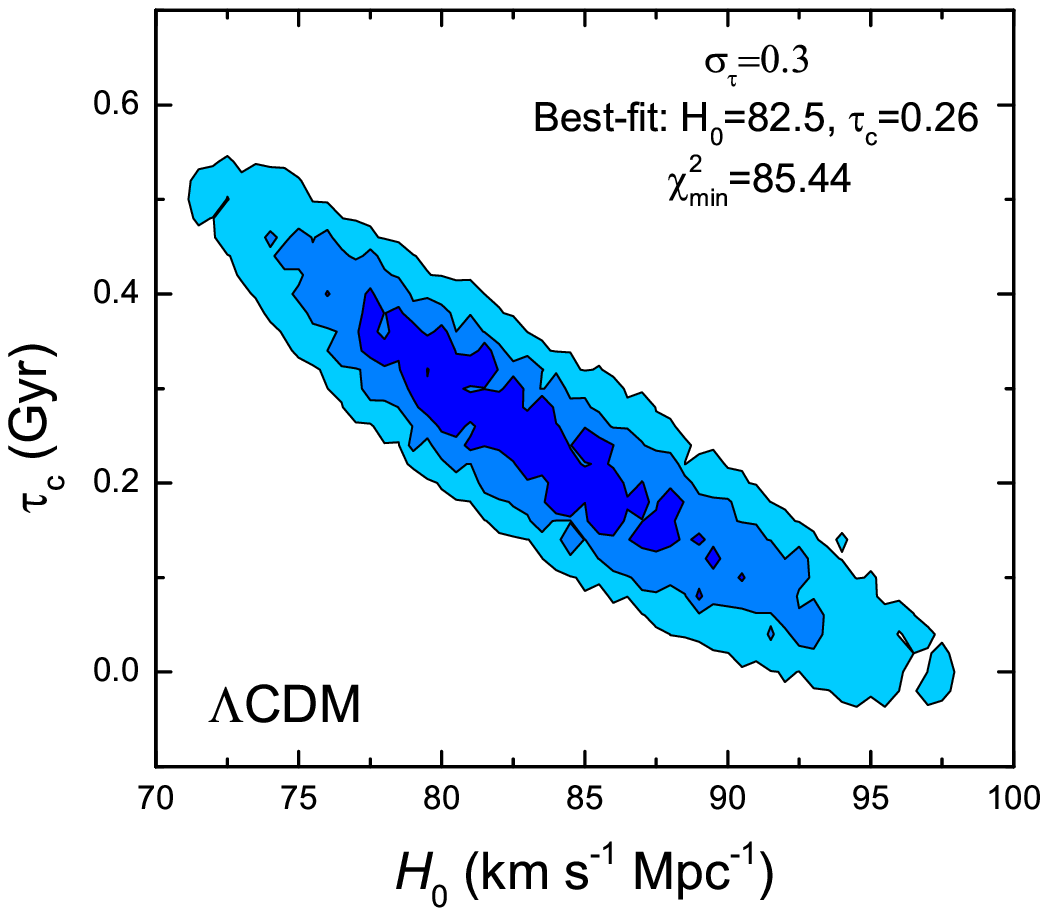}\hskip -0.9in
\includegraphics[angle=0,scale=0.7]{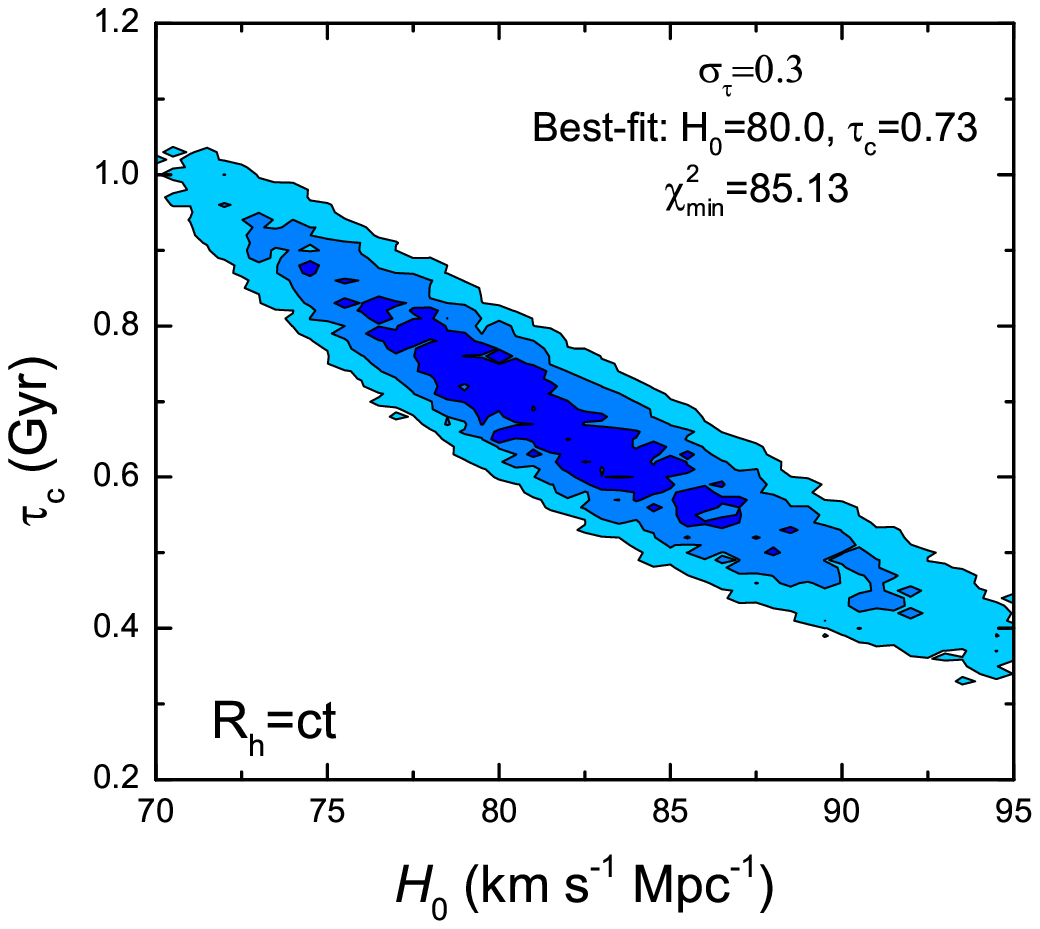}
    \caption{(a): $\Lambda$CDM with a Gaussian distribution for $\tau$ and a
    redshift dependent dispersion and mean value (see text). (b): Same as (a),
    except now for the $R_{\rm h}=ct$ Universe.}
\end{figure*}

An additional limitation of this type of work is the degree of uncertainty
in the galaxy-age measurement itself. It is difficult to precisely constrain
stellar ages for systems that are spatially resolved using stellar evolution
models. We may be grossly underestimating how uncertain the age measurements
of distant galaxies are. One ought to acknowledge this possibility and consider
its impact on cosmological inferences. For example, redoing our analysis
using a Gaussian $P(\tau)$ and an assumed dispersion $\sigma_{\tau}=0.3$,
but now with an additional $24\%$ uncertainty on the age measurements,
(i.e., twice as big as the value quoted in Dantas et al. 2009, 2011,
and Samushia et al. 2010), produces the results shown in Figure~13.
Panel (a) in this plot shows the corresponding distributions of
$\tau_{c}$ and $H_{0}$ in the concordance $\Lambda$CDM model, with best-fit
values $(H_{0}, \tau_{c})=(77.5, 0.86)$. Not surprisingly, a comparison of
Figure~13(a) with the left-middle panel in Figure~4 shows that,
as the uncertainty in the age measurement increases, the constraints
on model parameters weaken; nonetheless, the best-fit values of
$H_{0}$ and $\tau_{c}$ are more or less the same.

The comparison between $\Lambda$CDM (Figure~13a) and $R_{\rm h}=ct$
(Figure~13b) may be summarized as follows: the best-fit results are more
or less the same for both the $12\%$ and $24\%$ uncertainties. Based solely
on their minimum $\chi^{2}$ values, both models fit the passive galaxy
age-redshift relationship comparably well.
The AIC does not favour either $R_{\rm h}=ct$ or the concordance $\Lambda$CDM model,
regardless of how uncertain the galaxy-age measurements are, with relative
likelihoods of $\approx 51\%$ versus $\approx 49\%$.

\begin{figure*}[h]
\hskip -0.04in
\includegraphics[angle=0,scale=0.7]{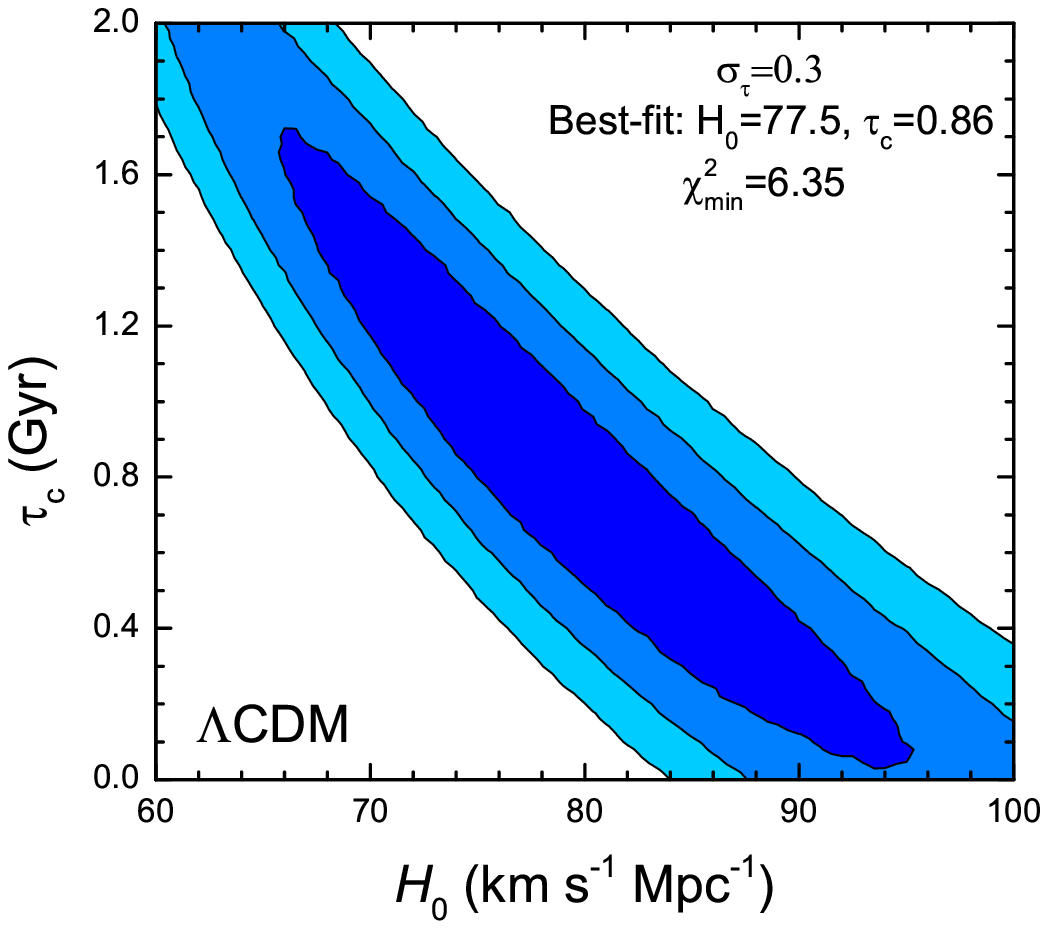}\hskip -0.9in
\includegraphics[angle=0,scale=0.7]{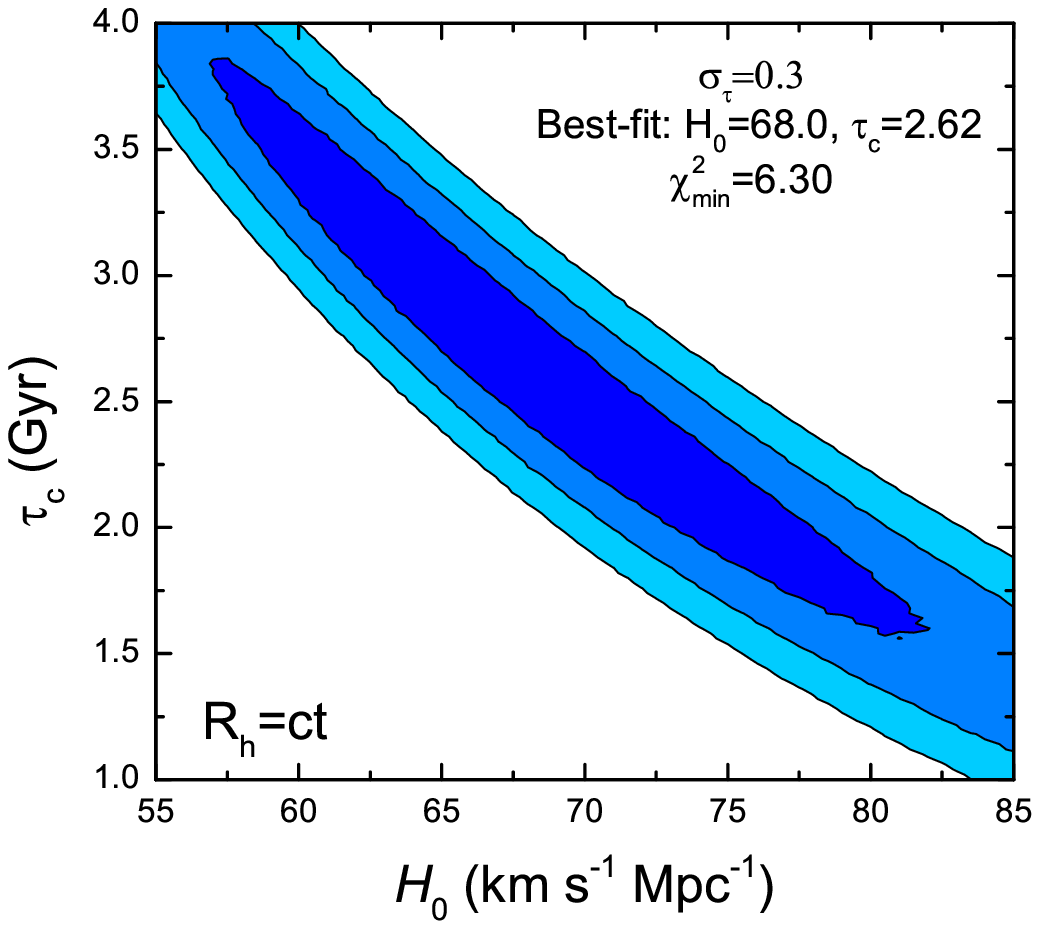}
    \caption{(a): $\Lambda$CDM with a Gaussian distribution of $\tau$
    and an (assumed) uncertainty of $24\%$ in the age measurement. (b): Same as (a),
    except now for the $R_{\rm h}=ct$ Universe.}
\end{figure*}

\vskip-0.2in
\acknowledgments
We are very grateful to the anonymous referee for providing a
thoughtful and helpful review, and for making several important
suggestions to improve the presentation in the manuscript.
This work is partially supported by the National Basic Research Program (``973" Program)
of China (Grants 2014CB845800 and 2013CB834900), the National Natural Science Foundation
of China (grants Nos. 11322328 and 11373068), the One-Hundred-Talents Program,
the Youth Innovation Promotion Association, and the Strategic Priority Research Program
``The Emergence of Cosmological Structures" (Grant No. XDB09000000) of
the Chinese Academy of Sciences, and the Natural Science Foundation of Jiangsu Province
(Grant No. BK2012890).
F.M. is grateful to Amherst College for its support through a John Woodruff Simpson
Lectureship, and to Purple Mountain Observatory in Nanjing, China, for its hospitality
while part of this work was being carried out. This work was partially supported by grant
2012T1J0011 from The Chinese Academy of Sciences Visiting Professorships for Senior
International Scientists, and grant GDJ20120491013 from the Chinese State Administration
of Foreign Experts Affairs.

\end{document}